\documentclass[aps,prb,showpacs,twocolumn,superscriptaddress]{revtex4}
\usepackage{bm,amssymb,amsmath,color,subfig}
\usepackage[pdftex]{graphicx}

\newcommand{\Eq}[1]{Eq.~\eqref{#1}}
\newcommand{\Eqs}[1]{Eqs.~\eqref{#1}}

\newcommand{\Fig}[1]{Fig.~\ref{#1}}
\newcommand{\Sec}[1]{Sec.~\ref{#1}}

\newcommand{\beq}{\begin{equation}}
\newcommand{\eeq}{\end{equation}}

\newcommand{\beqa}{\begin{eqnarray}}
\newcommand{\eeqa}{\end{eqnarray}}

\newcommand{\Beqa}{\begin{eqnarray*}}
\newcommand{\Eeqa}{\end{eqnarray*}}
\newcommand{\nn}{\nonumber}

\DeclareMathOperator{\sgn}{sgn}

\newcommand{\etal}{\textit{et al.}} 

\begin{document}

\title{Effects of the contacts on shot noise in graphene nano-ribbons}

\author{A.D. Wiener}
\author{M.  Kindermann}
\affiliation{School of Physics, Georgia Institute of Technology, Atlanta, Georgia 30332, USA}

\begin{abstract}
We investigate the shot noise of an impurity-free graphene flake as a function of the chemical potential. For large width to length ratios, this noise has been predicted and observed to exhibit universal characteristics at the Dirac point. Furthermore, a sharp decrease of the shot noise with increasing carrier density has been predicted. This decrease has also been observed in experiments, but with much smaller slope than predicted. We reconcile this discrepancy between theory and experiment by including the effects of the contacts to the graphene ribbon. \end{abstract}

\maketitle

\section{Introduction}
Impurity-free graphene\cite{novoselov:sci04,zhang:nat05,berger:jpc04,neto:rmp09} near its Dirac point has been predicted to exhibit non equilibrium current fluctuations, or shot noise.~\cite{tworzydlo:prl06,danneau:prl08,dicarlo:prl08} These fluctuations are counterintuitive, as shot noise vanishes in conductors without electron scattering.~\cite{khlus:jetp87,lesovik:jetp89,reznikov:prl95,kumar:prl96} They are due to evanescent waves that backscatter electrons, even in the absence of impurities.~\cite{tworzydlo:prl06}

Moreover, Tworzydlo \etal~\cite{tworzydlo:prl06} have predicted that the shot noise in a clean sheet of graphene at its Dirac point (zero chemical potential) has universal characteristics; the Fano factor $F$, defined as the shot noise normalized by the mean current and expressed in units of the electron charge $e$, takes the value $F=1/3$. This prediction has generated much theoretical~\cite{cayssol:prb09,rycerz:prb09,snyman:prb07,sanjose:prb07,lewenkopf:prb08,schuessler:prb09,sonin:prb08,golub:prb10,wiener:prb11}  and experimental~\cite{dicarlo:prl08,danneau:prl08} interest in the shot noise of graphene.

One goal of the ensuing experimental activity was to confirm the existence of shot noise due to evanescent waves in graphene.  A Fano factor $F\sim 1/3$ at the Dirac point of graphene has been observed, as reported in Refs.\ \onlinecite{dicarlo:prl08,danneau:prl08}, in agreement with the theory of Ref.\ ~\onlinecite{tworzydlo:prl06}.  This observation, however, does not provide unambiguous evidence of the mechanism discussed in Ref.~\onlinecite{tworzydlo:prl06}, since a Fano factor $F=1/3$ is expected not only in clean graphene at its Dirac point, but also in disordered metals. 

In the experiment of Danneau \etal,~\cite{danneau:prl08} the Fano factor reached a peak value $F\sim 1/3$ and displayed a strong dependence on the chemical potential. This provides further evidence for evanescent wave transport; as the chemical potential departs from the Dirac point, states that are evanescent at the Dirac point become propagating, which decreases backscattering and hence the Fano factor. This would not be the case if the measured shot noise was due to impurities. 
 
However, the energy scale of the measured dependence of the Fano factor on the chemical potential was considerably larger than theory predicts.~\cite{tworzydlo:prl06} As the dependence of $F$ on the chemical potential provides the main  experimental evidence for evanescent wave transport in graphene to date, this deviation from the theoretical prediction is disturbing.

In this article, we show that the measured dependence of the Fano factor on chemical potential is indeed consistent with the assumption that it originates from evanescent waves when the effects of the electrical contacts are taken into account. It has been shown by first-principles calculations that contacts to graphene have two main effects on electron transport: doping of the pieces of graphene underneath the contacts and electron scattering at the interface from contact to graphene. It turns out that the Ti-graphene contacts used in the experiment of Ref.~\onlinecite{danneau:prl08} are highly transparent. In this article, we therefore neglect contact scattering and focus on the effects of doping through the contacts. 

While the theory of Ref.\ ~\onlinecite{tworzydlo:prl06} assumes a constant electric potential on the graphene ribbon, the local doping of graphene by the contacts causes that potential to be space-dependent.  We show that this space-dependence  results in an increase of the voltage scale of $F$, as observed experimentally.  This effect can be understood in terms of an effective reduction of the length of the graphene ribbon to a region around the potential minimum.  Remarkably, the maximum of the Fano factor remains $F\approx1/3$ when the space-dependence of the potential caused by doping through the contacts is taken into account, also in agreement with the experiment of Danneau \etal~\cite{danneau:prl08} The reported calculation thus lends additional support to an interpretation of the experiment of Ref.\  \onlinecite{danneau:prl08} in terms of evanescent waves in graphene.

This article is organized as follows: after a description of the model in \Sec{mod}, we employ a conformal map technique in \Sec{potential} to calculate the electric potential on the graphene ribbon due to the contacts. We then consider the effects of screening of this potential by the electrons in the graphene flake and identify a regime where such screening is negligible.  In that regime, we then calculate the electron transmission through the graphene ribbon and the resulting Fano factor. We first do this analytically for the semiclassical regime in section \Sec{semiclassical}. In \Sec{numerical}, we then calculate the transmission from  numerically obtained wave functions, finding good agreement with both the semiclassical results in their regime of validity and experimental observations. We conclude with a summary in \Sec{conclusions}.

\section{Model}\label{mod}

\begin{figure}
\centering
\begin{tabular}{c}
\begin{tabular}{cc}
\subfloat[ ]{
  \includegraphics[width=0.5\columnwidth]{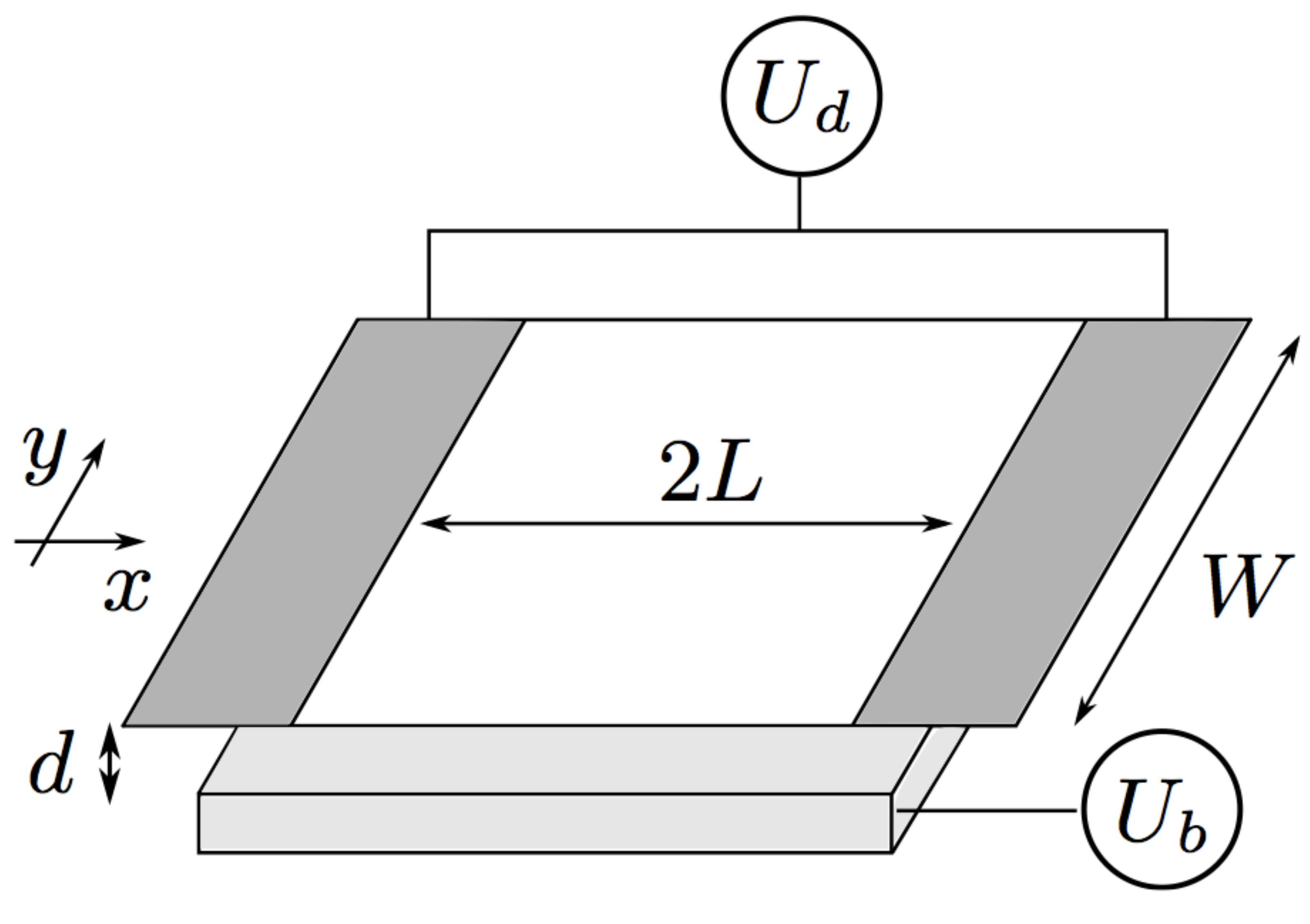}
  \label{fig:ribbon}
 }
 &
\subfloat[ ]{    
   \includegraphics[width=0.45\columnwidth]{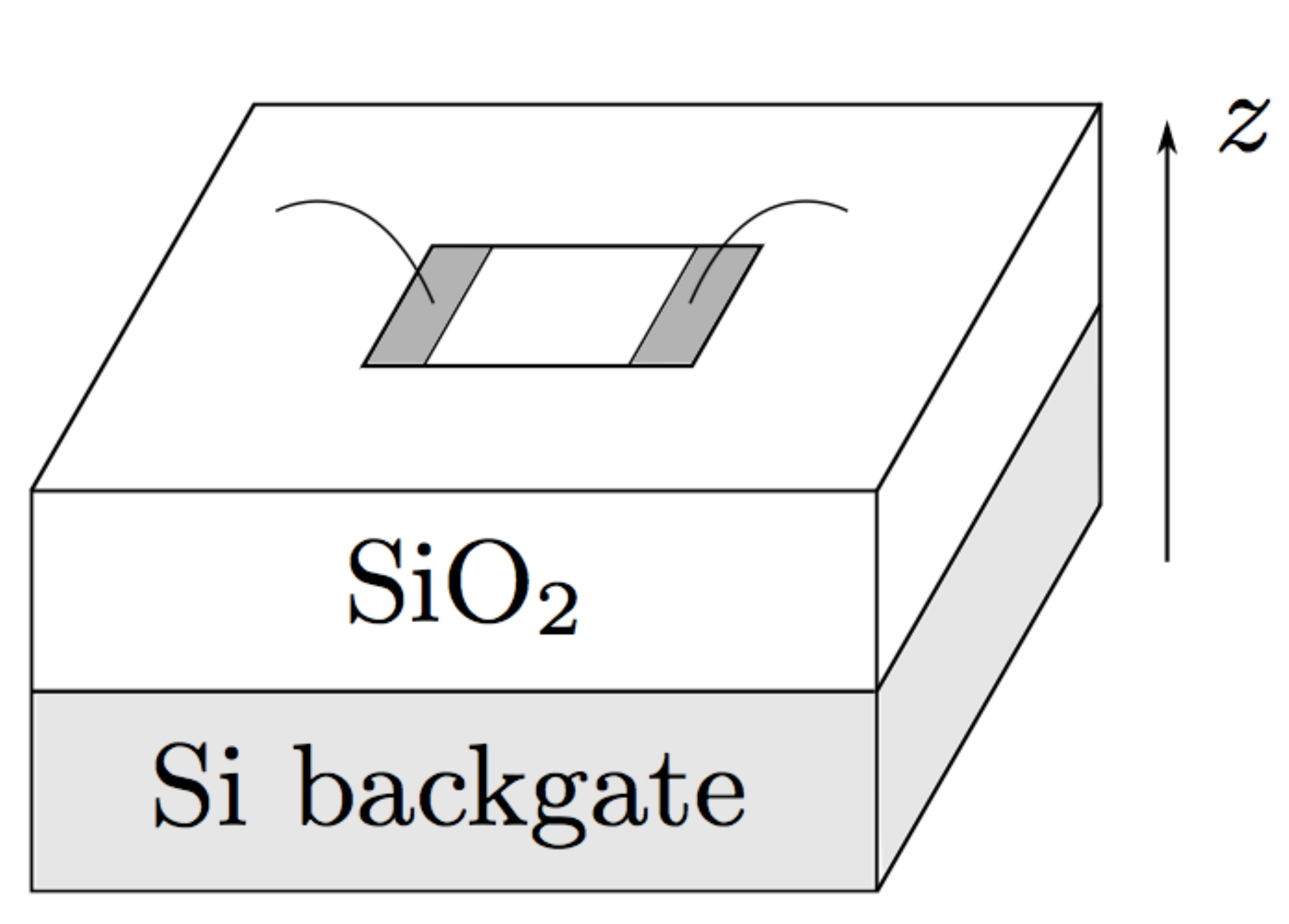}
   \label{fig:sample}
 } 
 \end{tabular}
  \\
 \subfloat[ ]{
  \includegraphics[width=\columnwidth]{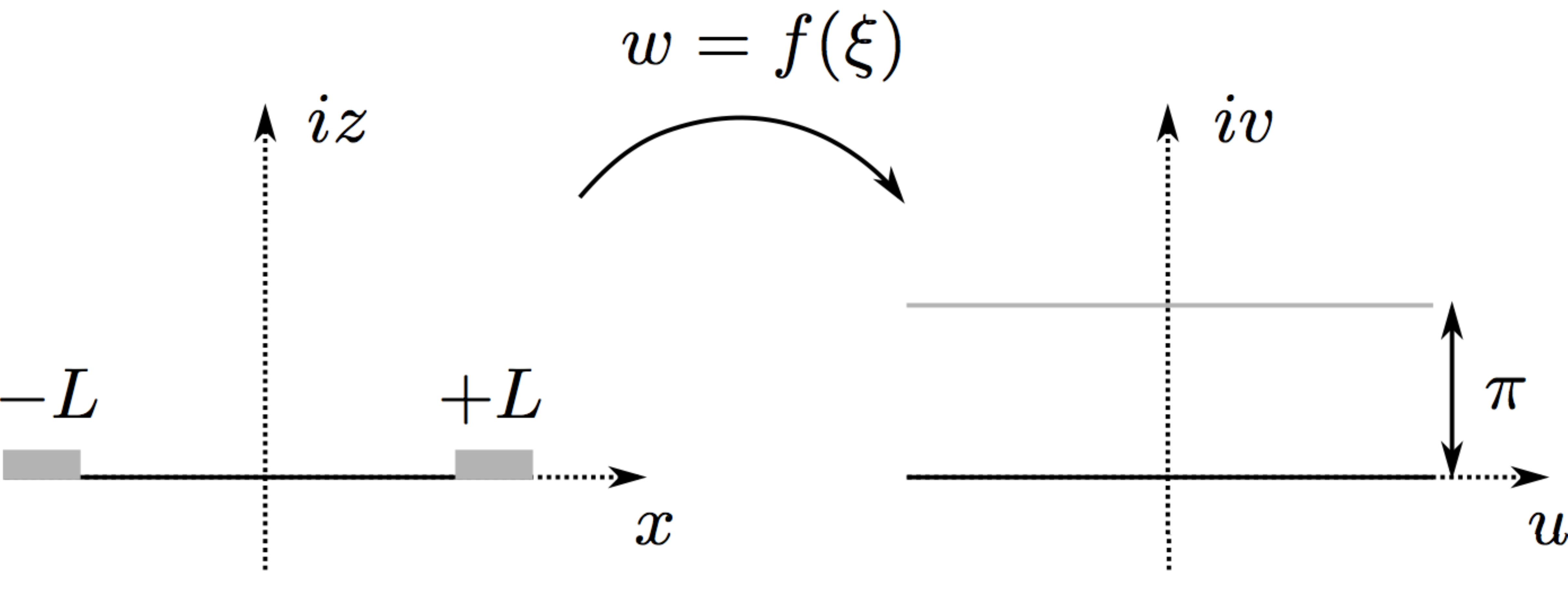}
   \label{fig:map}
   }
 \end{tabular}
\caption{(a) Graphene nano-ribbon geometry consisting of a graphene sheet with highly doped lead regions (dark gray rectangles) separated by an undoped region (white rectangle). An insulating SiO$_2$ layer of thickness $d$ (not shown) separates the graphene sheet from a Si backgate (light gray rectangle). (b) An example of a simplified experimental geometry as considered in Ref.~\onlinecite{danneau:prl08}. (c) For the $d=0$ case, the conformal map $w=f(\xi )$ maps the $(x,iz)$ plane to the $(u,iv)$ plane. The undoped graphene regions are sent to the real line $v=0$ while the doped graphene regions are sent to the line $v=\pi$. The height of the doped lead regions shown here is for illustration purposes and has no physical meaning.}
\label{fig:schematics}
\end{figure}

In the experiment of Danneau \etal,~\cite{danneau:prl08} a graphene nano ribbon (GNR) was mechanically exfoliated, deposited on a Si/SiO$_2$ substrate as in \Fig{fig:sample}, and brought in close proximity with Ti contacts. An insulating SiO$_2$ layer of thickness $d$ separates the GNR from a Si backgate.  We assume that the electronic structure of the sections of the ribbon underneath the contacts is unmodified, except for a doping through the contacts. It has been shown using density-functional calculations that this model is a good approximation for non-wetting contacts~\cite{barraza:prl10} and also for (wetting) Ti contacts at low energies.~\cite{barraza:nano12}

To find the electric potential $U$ on the GNR, we model the system by two metal half-planes separated by a distance $2L$, representing the contacts, and a third metallic plane, representing the Si backgate, at distance $d$ below the GNR, as shown in \Fig{fig:ribbon}. All metals are assumed to be perfect, and hence equipotentials, and we choose coordinates such that the contact edges are along the $y$-direction and  at $x=\pm L$.  The $z$ axis is  perpendicular to   backgate and GNR.  

In equilibrium, the contact plates are at identical electric potentials $U_d$, while the backgate has a potential $U_b$. Under the reasonable assumption that the distance between the contacts and the GNR is negligible, $U_d$ is also the electrostatic potential in the graphene regions underneath the contacts. In the absence of a backgate and in the limits that we take below, the potential on the ribbon close to the center of the GNR differs from $U_d$  by the work function difference between the GNR and the metal contacts. The backgate is used to manipulate the potential in this region where, in the same limits, all electron scattering occurs.

The low-energy Dirac model of the electron dynamics in the GNR   ($\hbar=1$) is,
\beq \label{eq:Hamiltonian}
H_\gamma= v \boldsymbol{\sigma}_\gamma \cdot  \boldsymbol{p} + V(x),
\eeq
where $\boldsymbol{\sigma}_\gamma=(\gamma\sigma_x,\sigma_y)$ is a vector of Pauli matrices with the valley index $\gamma = \pm 1$, $\boldsymbol{p}=-i\boldsymbol{\nabla}$ is the electron momentum and  $v$ is the electron velocity. The potential energy of electrons on the GNR is given in terms of the electric potential $U$ by  $V(x)=-eU(x)$. 
Using translational invariance in the $\hat{y}$-direction and choosing boundary conditions that do not mix transverse modes,\cite{berry:prsla87, brey:prb06} the momentum along the $\hat{y}$-direction is conserved, and the scattering problem for wave functions $\psi_q$ with $\hat{y}$-momentum $q$ at energy $\varepsilon$ becomes one dimensional:
\beq \label{Dirac}
 v (-i\gamma\sigma_x \partial_x +q \sigma_y)\psi_q=\left[\varepsilon-  V(x)\right]\psi_q.
\eeq
 
We focus on the most interesting regime of the Fano factor measurement of Danneau \etal,~\cite{danneau:prl08} namely an interval   of chemical potentials $-V(0)$ in the middle of the GNR with width $\Delta \mu $. The Fano factor reaches its maximum $F\approx 1/3$ at the center of this interval and decays monotonically to a value $F<0.25$ at the boundaries. Existing theory~\cite{tworzydlo:prl06} assumes a piecewise constant shape of $V$, with $V=V_\infty \gg v/L$ underneath the contacts and $V=V_0$ in between. This potential shape is obtained from our electrostatic model in the limit $d\to0$, with $V(0)=V_0=-eU_b$. 

Such theory predicts oscillations of the Fano factor of period $\Delta \mu \simeq v/L$ (setting $\hbar=1$). In the experiment,~\cite{danneau:prl08} however, the interval $\Delta \mu$ was found to be at least a factor of $2$ larger than this prediction. Of course, in the experiment $d \neq 0$, and the electrostatic model of Ref.\ \onlinecite{tworzydlo:prl06} is oversimplified. Below, we analyze the effects of nonzero distance $d$ between the GNR and backgate, assuming that $d$ is small in a sense that will be specified. In that case, one still has $V(0) \approx -eU_b$, but the potential $V$ is no longer piecewise constant. We  show below that such non-zero $d$ resolves the discrepancy between theory and experiment.

\section{Calculation of the potential} \label{potential} 
\subsection{The conformal map}
We start with the calculation of the electrostatic potential $U(x)$ on the GNR. In a first approximation, we neglect doping of the GNR, effectively assuming it is a perfect insulator. We will justify this approximation for small $d$ in the following section. Additionally assuming that the contacts lie directly on top of the GNR and that the distance $d$ to the backgate is small compared to the length $L$ of the ribbon, $d\ll L$, we find the potential $U$ on the GNR to a good approximation (at distance $\Delta x\gg d$ from the contacts) from the electric field ${\bf E}_0$ on the ribbon when $d=0$:
\beq \label{Ed}
U_0=U_b+\hat{z}\cdot{\bf E }_0d/\epsilon_r,
\eeq
 where $\epsilon_r=\epsilon/\epsilon_0 \approx 3.9$~\cite{gray:09} is the relative electric permittivity of the $\mathrm{SiO_2}$ layer separating the GNR from the backgate. Here, $\epsilon_0$ is the vacuum permittivity and $\epsilon$ is the permittivity of silicon oxide.

The electric field ${\bf E}_0$ at $d=0$ may be found by an appropriate conformal mapping, exploiting translational invariance in the direction along the contact edges.  The electric potential $U$ satisfies the two-dimensional Laplace equation $\nabla^2 U(x,z)=0$, along with the boundary conditions
\beqa
\label{eq:potBC}
U(\left| x\right| >L,z=0) &=& U_d, \nn \\
U(\left| x\right| <L,z=0) &=& U_b.
\eeqa
The conformal map
\beq
f(\xi ) = \ln\left(\frac{\xi -L}{\xi +L}\right)
\label{eq:map}
\eeq
from the complex plane $\xi=(x,iz)$ into the complex plane $w=(u,iv)$ maps these boundary conditions onto those of a parallel plate capacitor, with a plate separation of $\pi$,  as shown in \Fig{fig:map}. When $z\to 0$, the map $f(\xi )$ sends the graphene regions under the contacts at $|x|>L$ in the $\xi$-complex plane to the real line $v=0$ in the $w$-complex plane. Similarly, for $z\to 0$, the region $|x|<L$ in the $\xi$-plane is mapped to the line $v=\pi$ in the $w$-plane.  

The potential in the $(u,iv)$ plane is thus
\beq
U(u,v)=\frac{(U_b-U_d)v}{\pi}.
\eeq
Under the map \Eq{eq:map} this gives the electric potential  
\beq
U(x,z)=\frac{U_b-U_d}{2\pi}\tan^{-1}{\left(\frac{2zL}{x^2+z^2-L^2}\right)}+U_b.
\label{eq:fullpot}
\eeq
in the original $(x,iz)$ plane.

By differentiation, we now find the electric field $\bf{E}_0$ on the GNR for $d=0$ from \Eq{eq:fullpot} and, using \Eq{Ed}, we obtain the electric potential on the ribbon at $d \ll L$ and $L-|x| \gg d$ as  
\begin{align} \label{pot}
U_0(x)&=U_b+\hat{z}\cdot{\bf E }_0(x,0) d/\epsilon_r\nn\\
&= U_b+\frac{U_b-U_d}{\epsilon_r\pi\left(1-x^2/L^2\right)}\left(\frac{d}{ L}\right).
\end{align}

\subsection{Corrections due to screening by the graphene sheet}\label{screening}

Next, we account for the screening of the electric potential generated by the contacts arising from the electrons in the ribbon itself. Quantifying this screening requires a determination of the electron density $n(x)$ that accumulates on the ribbon due to the potential ${V}(x)$. The corresponding charge density  $-en$  induces a screening electric field ${\bf E}_{sc}$, which in turn modifies the   potential on the ribbon $U$. Under our assumption $d\ll L$ we have  $U(x) = U_b+ \hat{z}\cdot[{\bf E }_0(x,0)+ {\bf E}_{sc}(x,0) ] d/\epsilon_r$, and ${\bf E}_{sc}$ is approximately given by Gauss' law: 
\beq \label{Esc}
\hat{z}\cdot {\bf E}_{sc}(x,0) = -\frac{en(x)}{\epsilon_0}.
\eeq
  
A determination of the electron density at zero temperature and chemical potential
\beq \label{n}
n(x)=\sum_{q,\varepsilon<0} |\psi_{q,\varepsilon}(x)|^2
\eeq
requires the wave functions $\psi_{q,\varepsilon}(x)$. Rather than carrying out the requisite quantum mechanical calculation,  we identify a parameter regime where the screening field $E_{sc}$ may be neglected, $|E_{sc}| \ll E_0$. The key observation allowing this approximation is that at $U_b=0$, $V(x)$ is of first order in $d$. Semiclassically, the induced electron density is of order $V^2/v^2 \sim d^2$. One thus expects that  $E_{sc} ={\cal O}(d^2)$, while $E_{0} ={\cal O}(d^0)$, such that $|E_{sc}| \ll E_0$ at $d\to 0$. Based on this semiclassical reasoning, one thus expects that for $d$ below a critical distance $d_c$ the effects of screening may be neglected. In appendix \ref{appA}, we rigorously establish the existence of such a critical distance for the relevant interval  $\Delta \mu$ of gate voltages and we compute $d_c$. We find
 \beq \label{dc}
d_c=\epsilon_r v\hbar (\epsilon_0 L )^{3/4}e^{-7/4}  (U_b-U_d)^{-1/4} .
\eeq
For the remainder of this article we assume that $d \ll d_c \ll L$. In that regime we have, to a good approximation, $V(x) =-e U_0(x)$ and $V(0) =- eU_b$. In the experiment of Danneau \etal~\cite{danneau:prl08} $d_c \approx 15 \,{\rm nm}$, while $d\approx 300 \,{\rm nm}$ and $L \approx 200\, {\rm nm}$. The experiment therefore is not in the limit that we assume. Our calculation therefore merely highlights the qualitative physics of the voltage scale enhancement observed in that experiment.

\section{Transport calculations} \label{semiclassical}
\subsection{The transfer matrix}
In order to calculate the $q$-dependent transmission probabilities for transport through a GNR as in \Fig{fig:ribbon}, we employ the transfer matrix method.~\cite{falko:prb06}  Without restriction, we fix the valley index $\gamma=+1$ in the Hamiltonian \Eq{Dirac}, and we confine our analysis to equilibrium at electrochemical potential $\mu=0$, such that we require the transfer matrix only for electrons with energy $\varepsilon=0$.

The transfer matrix $M(x,x')$ for the requisite Dirac spinors of a mode with transverse momentum $q$ and energy $\varepsilon=0$ satisfies the equation
\beq
i\partial_x M(x,x') = \left[iq\sigma_z +\frac{V(x)}{v}\sigma_x\right] M(x,x').
\label{eq:DiracTransfer}
\eeq
Additionally, $M$ satisfies the conditions~\cite{falko:prb06} $M(x,x)=I$, $M(x,x')=M(x,x'')M(x'',x')$, $\det M(x,x')=1$ and $M^\dagger (x,x')\sigma_xM(x,x')=\sigma_x$. The latter condition ensures current conservation. 

In order to extract the transmission probability from the transfer matrix, it is necessary to factor out the asymptotic evolution of the electron states at $|x|\to \infty$. This is accomplished by using matrices $A_{\pm}(x)$ that satisfy \Eq{eq:DiracTransfer} in the  regions under the contacts, where the potential is constant, $V = eU_d \equiv vk_{\rm F}$. The solution of this equation gives the $A_{\pm}$ matrices as
\beq \label{A}
A_{\pm}(x)=\sqrt{\frac{k_{\rm F}}{2p_x}}\left(\begin{array}{cc} \frac{p_x\pm iq}{k_{\rm F}}e^{\mp ip_xx} & \frac{-p_x\pm iq}{k_{\rm F}}e^{\pm ip_xx} \\ e^{\mp ip_xx} & e^{\pm ip_xx}\end{array}\right),
\eeq
where $p_x=\sqrt{k_{\rm F}^2-q^2}$. The columns of $A_{\pm}$ are made of right and left-moving states that are normalized to carry unit current. The transmission probability is extracted from the transfer matrix $M(x,y)$ as $T=1/|\alpha|^2$, where
\beq
\label{eq:Mextraction}
\left(\begin{array}{cc}\alpha & \beta^* \\ \beta & \alpha^*\end{array}\right) = \lim_{x\rightarrow\infty}A_+^{-1}(x)M(x,-x)A_-(-x).
\eeq
Here, $|\alpha|^2-|\beta|^2=1$ due to the current conservation condition.

\subsection{Analytic calculation of transmission probability}
\label{analytical}
In parts of the central graphene region with potential \Eq{pot}, transport is semiclassical and the transfer matrix can be found using an adiabatic approximation. This is the case whenever~\cite{falko:prb06} $|q V'/V (V^2/v^2-q^2)| \ll1$. Loosely speaking, this condition is met where the potential is large, such as near the contacts. It has been shown in Ref.\ \onlinecite{falko:prb06} that no electron scattering takes place in those regions. All shot noise is therefore produced in the regions that do not allow such an approximation, near the classical turning points where $V=\pm vq$.

As previously mentioned, we assume that the regions of interest, where electron scattering occurs, are at $|x|\ll L$. The precise form of the potential at $|x| \simeq L$ is thus irrelevant, and we may approximate the potential \Eq{pot} quadratically,~\footnote{Here we assume $\rho>0$. In reality, Ti contacts n-dope graphene, such that $\rho<0$.~\cite{barraza:nano12} The sign of $\rho$, however, can be changed by a particle-hole transformation $U=\sigma_z$, and is thus straightforwardly accounted for in our calculation.}

\beq
\label{quadpot}
V(x) =  \rho x^2- V_b,
\eeq
where  $\rho =e (U_d-U_b)d /\epsilon_r \pi L^3$ and $V_b=eU_b$ is set by the backgate voltage $U_b$. For a dimensional analysis, we first write the Dirac equation (\ref{Dirac}) with potential (\ref{quadpot}) at $q=0$  in terms of the dimensionless variable $\gamma = x/\tilde{x}$, where $\tilde{x}= (\rho/v)^{-1/3}$. This results in an energy scale  $ \Delta \mu \approx v/\tilde{x}$ for the interval $I_D$ of the first oscillation of $F$. 

To make rigorous analytical progress, we assume $|V_b| \gg v/\tilde{x}$. While this is not the most relevant limit experimentally, this calculation will provide physical insight into the transport problem. Our analytical approach decomposes the GNR into ``adiabatic regions,'' where the semiclassical approximation may be applied, and ``non-adiabatic regions'' near the classical turning points $V=\pm vq$ where it cannot. One finds that for $|V_b| \gg v/\tilde{x}$,  each non-adiabatic region is short enough for the potential to allow linearization throughout the region. The transfer matrix for a linear potential has been found exactly in Ref.\ \onlinecite{falko:prb06}. This, together with the adiabatic solution for the remaining regions and the composition rule $M(x,x')=M(x,x'')M(x'',x')$, allows us to construct the transfer matrix through the entire GNR.

We find the above condition $|V_b| \gg v/\tilde{x}$ for applicability of the described analytical approach by self-consistently assuming that the potential $V$ may be linearized in the non adiabatic regions. It has been shown~\cite{falko:prb06} that, in this case, transport through a non-adiabatic region around a pair of classical turning points $V=\pm vq$ is exponentially suppressed by a factor $\exp(-\pi v q^2/V')$. Thus, only modes with $q \lesssim \sqrt{V'/v}$  contribute significantly to transport, and we may neglect all other modes.  The condition quoted above for adiabatic electron dynamics thus effectively becomes $|vV'| \ll V^2$. Using the explicit form \Eq{quadpot} of the potential $V$, we find that this condition is fulfilled everywhere except in regions of length $\Delta x$ around the points $x_0$ with $V(x_0)=0$ that are short enough to allow linearization of $V$, that is $\Delta x \ll x_0$.  Therefore, in the above limit $|V_b|\gg v/\tilde{x}$ the transfer matrix can indeed be constructed from that of electrons in a linear potential and the one for adiabatic evolution.
 In appendix \ref{appB} we calculate transport through the GNR in this limit.

\subsection{Analytic results}
From the transfer matrix  \Eq{eq:PositiveTransfer} obtained in appendix \ref{appB}, one analytically extracts the transmission probability using \Eq{eq:Mextraction}. The resulting transmission probability takes the form
\begin{align}
\label{eq:AnalyticResult}
T &= \left|\alpha^2+ie^{-2i \phi}\left(b^*\right)^2\right|^{-2}, \nn \\
\phi &= \int_{\ell-x_0}^{x_0-\ell}\sqrt{V^2(x)/v^2-q^2}dx,
\end{align}
where $b=-\sqrt{2\pi}e^{\pi\theta /2}\theta^{1/2-i\theta}/\Gamma{\left(1-i\theta\right)}$ and $V(\pm x_0)=0$. The wave function acquires the phase $\phi$ from traversing the central adiabatic region separating the turning point at $x=\ell-x_0$ from the one at $x=x_0-\ell$.

\begin{figure}
\centering
\includegraphics[width=\columnwidth]{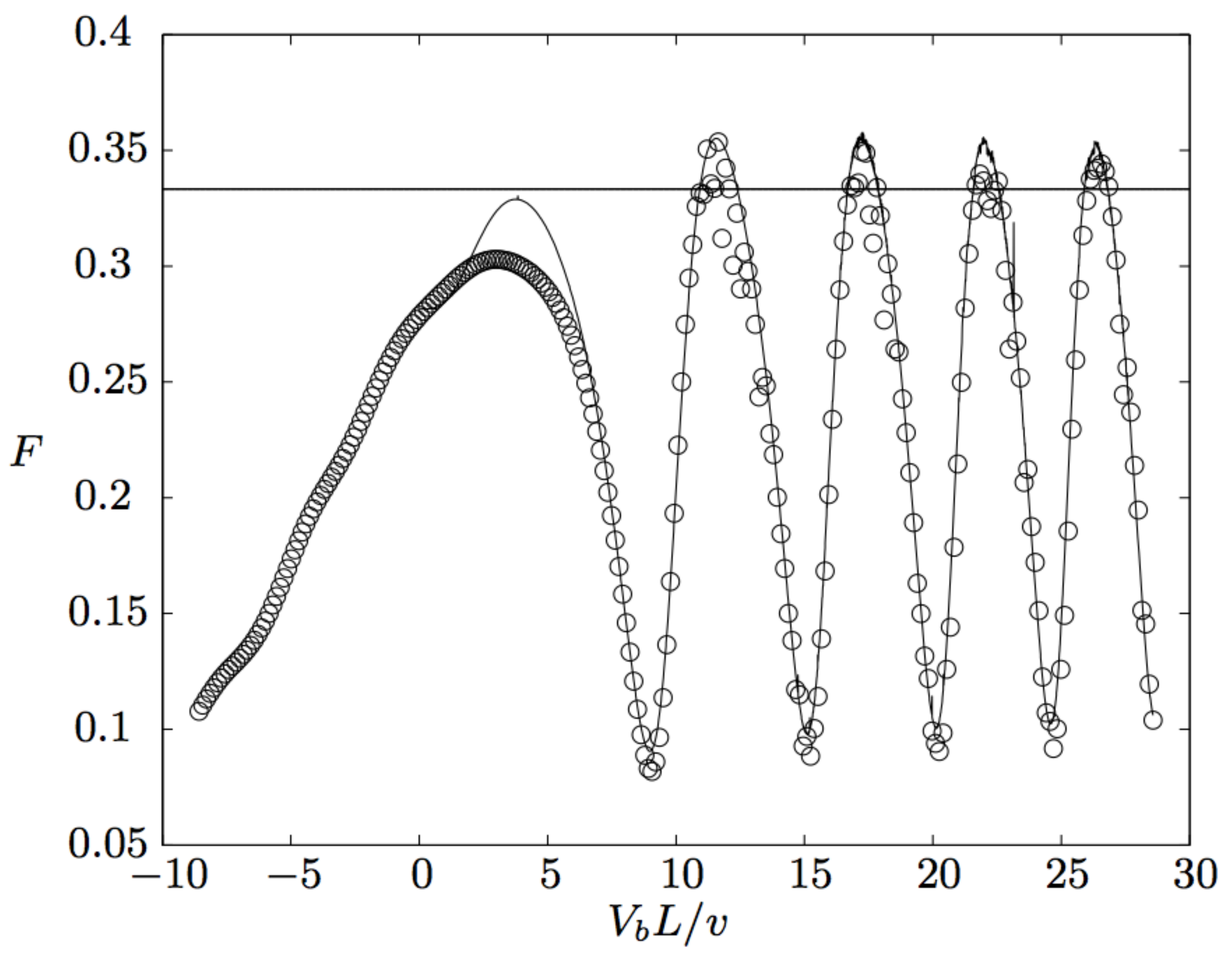}
\caption{Analytic (solid curve) and numerical (circles) results for the Fano factor $F$ as a function of backgate voltage in units of $v/L$, along with the constant $F=1/3$. The curves are generated assuming \mbox{$\hbar v=0.7 \times 10^{-9}$ $\mathrm{eV\cdot m}$}, \mbox{$L=500$ $\mathrm{nm}$} and \mbox{$\rho=10^{-6}$ $\mathrm{eV/m^2}$}. The analytic results are calculated from the transmission probability of \Eq{eq:AnalyticResult}. The numerical result is calculated from the numerically integrated wave functions according to \Eqs{Dirac}. The results differ at low backgate voltages where the turning points no longer lie within the linear region of the potential and the semiclassical approximation used here breaks down.}
\label{fig:fano}
\end{figure}

The Fano factor is found from the transmission probabilities $T_n=T(q_n)$ of modes with wavenumbers $q_n$ according to \Eq{eq:AnalyticResult} as \cite{lesovik:jetp89, buettiker:prl90}
\beq
\label{eq:FanoDiscrete}
F = \frac{\sum_n T_n\left( 1-T_n\right)}{\sum_nT_n}.
\eeq
In the continuum limit $W\gg L$, the sums over the mode index become integrals over the momentum $q$. The Fano factor is plotted as a function of the backgate voltage in units of $v/L$ in \Fig{fig:fano} (solid curve). The curve shows oscillations in gate voltage with a maximum of $F\sim 1/3$, as predicted by the theory~\cite{tworzydlo:prl06} describing evanescent mode transport in a piecewise constant potential. However, the width of the peaks is broader than predicted by the theory of Ref.\ \onlinecite{tworzydlo:prl06}, in agreement with the observations of Danneau \etal~\cite{danneau:prl08}

Stretching the limits of applicability of our semiclassical approach, \Eq{eq:AnalyticResult} predicts that the first oscillation of the Fano factor with backgate voltage has period $\Delta \mu= (3\pi/2)^{2/3}v/\tilde{x}$. Substituting $\rho =e (U_d-U_b)d /\epsilon_r\pi L^3$, the predicted oscillation period 
\beq
\Delta \mu= \eta \Delta \mu|_{d=0}
\eeq
is enhanced by a factor $\eta=(9 e |U_b-U_d|d/4\epsilon_rv\pi^2)^{1/3}$ compared to the period $\Delta \mu|_{d=0}= \pi v/L$ for a piecewise constant potential. The enhancement factor for the parameters of the experiment of Ref. \onlinecite{danneau:prl08} takes a value of $\eta \approx 2$. The physical reason for this enhancement is now clear; rather than being confined to the ribbon of length $2L$, the electron states in the presence of doping from the contacts form standing waves between the two non adiabatic regions that introduce electron scattering around $x=\pm x_0$. The width $\Delta \mu$ of the first  oscillation    of $F$ as a function of backgate voltage is thus enhanced. 

Unfortunately, our above analytic approach breaks down at $|V_b| \lesssim v/\tilde{x}$, the regime $V_b\in I_D$ of the experimentally most interesting first Fano factor oscillation, and the above considerations are not quantitatively correct. Physically, this is the regime where the potential may not be linearized in the non adiabatic regions. In this case, the two non adiabatic regions merge into one. To quantitatively access that regime, we next perform numerical calculations of the transfer matrix. 

\subsection{Numerical calculation of the transmission probability} \label{numerical}
In this section, we obtain the transmission probability by numerical integration of \Eq{Dirac}.  
Numerical results for the Fano factor  are plotted in \Fig{fig:fano} (circles). At $|V_b| \gg v/\tilde{x}$, where our analytic approach is justified, the curve agrees with our analytic results, as expected.  Our numerical calculation confirms what we had observed analytically for the first and most relevant Fano factor maximum with $|V_b| \gg v/\tilde{x}$. Doping from the contacts increases the voltage scale of the Fano factor oscillations. From the full width at half maximum of our numerical results, we conclude that the width of the first peak of the Fano factor is enhanced by a factor $\eta \approx 1.7$ compared to the theory neglecting doping from the leads. 

Note also, that, as claimed in the introduction, the maximal value of the Fano factor is $F \approx 1/3$ as for the idealized, piecewise constant potential assumed in Ref.\  \onlinecite{tworzydlo:prl06}. Comparing our numerical result with the Fano factor extracted experimentally from the voltage-dependence of the current fluctuations in Ref.\    \onlinecite{danneau:prl08}, we note that the numerically determined maximum $F\approx 0.305$ is as compatible  with the experimental value $F=0.318$  as the value  $F=1/3$ predicted using the idealized potential of Ref.\  \onlinecite{tworzydlo:prl06}.

 Finally we comment on the Fano factor oscillations at large $V_b$ in  \Fig{fig:fano}. They do not show in the experimental data,  Fig.\  3 of Ref.\  \onlinecite{danneau:prl08}. This discrepancy could have various reasons. We here briefly speculate on two of them. First,   the gate voltage scale where the oscillations in \Fig{fig:fano} set in is about $V_b \approx 50 {\rm meV}$ (the ribbon in the experiment has length $L=200 {\rm nm}$). This is beyond the range of gate voltages where our quadratic approximation Eq.\ (\ref{quadpot}) is justified for the work function difference $|U_d -U_b|\approx 200 {\rm meV}$ between graphene and the Titanium contacts that were used in the experiment of Ref.\ \onlinecite{danneau:prl08}. Our theory thus really only reliably describes the central Fano factor maximum of the experiment of Ref.\  \onlinecite{danneau:prl08}.  The Fano factor oscillations at $V_b L/v >10$ shown in  \Fig{fig:fano} are outside the regime of validity of our theory. Second, we note that our prediction   \Fig{fig:fano} assumes a perfect geometry of a ribbon between two perfecty straight and parallel contact edges. In reality, of course, this is not the case and the length of the ribbon varies over its width, in particular in samples with a large $W/L$-ratio such as the ones that were measured for Fig.\  3 of Ref.\  \onlinecite{danneau:prl08}. Such variations will wash out the predicted oscillations. It will suppress them the more the further the energy is from the Dirac point.


\section{Discussion and Conclusions} \label{conclusions}
In conclusion, we have shown that the anomalously large voltage scale observed in the GNR shot noise experiments of Danneau \etal~\cite{danneau:prl08} is consistent with evanescent wave transport when the effect of doping by the contacts is accounted for. We have identified a regime of small graphene-backgate distances where the effects of screening by conduction electrons in the GNR can be neglected.  While not well satisfied in the experiment of Ref.\ \onlinecite{danneau:prl08},  this limit gives insight into the qualitative physics of the voltage scale enhancement observed by  Danneau \etal~\cite{danneau:prl08}

In this regime, we find the electric potential on the GNR with contacts, and we use it to obtain the Fano factor as a function of the backgate voltage. We employ both a semiclassical and a numerical approach. The semiclassical approach illuminates the origin of the predicted increased gate voltage period of the Fano factor. The potential due to doping from the leads introduces electron scattering around, generally, two pairs of classical turning points of the conduction electrons at distance $2 x_0< 2L$. The standing waves that form in between cause oscillations of the Fano factor with period $\Delta \mu \simeq v/x_0$, which is larger than the scale $\Delta \mu \simeq v/L$ of the same oscillations without doping by the contacts. Our numerical results show that the contact potential enhances $\Delta \mu $ by a factor of $\eta\approx 2$, consistent with the experimental observations.~\cite{danneau:prl08} Our calculations   demonstrate that an interpretation of the experiment by  Danneau \etal~\cite{danneau:prl08} in terms of evanescent waves is possible and strongly indicated, despite the discussed discrepancy with the original theory.

 \section{Acknowledgement}
 
 We acknowledge support by the NSF under DMR-1055799 and DMR-0820382.

\section{Appendix A} \label{appA}
In this appendix, we identify a parameter regime where the screening field $E_{sc}$ due to doping of the GNR may be neglected in our calculation of the Fano factor $F$. Self-consistently, we thus assume that $V(x)=-e U_0(x)$  for the argument below. 
In our limits only the region $|x| \ll L$ contributes to the shot noise (see main text) and we may approximate the potential $V$ as in Eq.\ (\ref{quadpot}). As explained in the main text, the central length scale in the problem then is $\tilde{x}= (\rho/v)^{-1/3}$, the typical wavelength of the electron states at the Fano factor maximum, which sets the energy scale $ v/\tilde{x}$ of the interval $\Delta\mu$ of the first oscillation of $F$. 

We first show that transport states with transverse momenta  $q \gg 1/\tilde{x}$ are irrelevant for the determination of $F$. To see this, we perform an adiabatic expansion of Eq.\ (\ref{Dirac}) in the spirit of  Ref.~\onlinecite{falko:prb06}, but for $vq > V(x)$. That expansion is valid if $|V'(x)| \ll vq^2$, which is fulfilled at $|x|<\tilde{x}$ for the transport states (that is, states at the Fermi level, with energy zero) with $q \gg 1/\tilde{x}$ in the window of backgate voltages $\Delta \mu$  of interest. For such $q$, this adiabatic calculation  results in a transmission   of electrons through the region $|x|<\tilde{x}$, which is  exponentially suppressed in $q\tilde{x} \gg 1$. Consequently  $T_n \ll 1$, and states with  $q\gg 1/\tilde{x}$ do not contribute to the Fano factor, Eq.\ (\ref{eq:FanoDiscrete}). 
 
Moreover, $F$, Eq.\ (\ref{eq:FanoDiscrete}), depends only on the transmission eigenvalues $T_n$.  These eigenvalues receive no contributions from regions in space with semiclassical  electron dynamics,  which is explicitly evident in section \ref{analytical}. According to  Ref.~\onlinecite{falko:prb06}, semiclassical dynamics takes place for $|q V'/V k^2|\ll1$ (here $V'$ denotes the first derivative of $V$), and regions  where that condition is met may therefore be disregarded in our calculation.~\footnote{In our above electrostatic calculation, the electric field $E_0$ diverges at the edge of the contacts when $d\to 0$, which one may suspect could introduce electron scattering. We note, however, that this is an artifact of our approximation that expands $U$ to first order in $d$. Moreover, the effects of the large electric fields that do emerge at the edges of metal contacts to graphene were taken into account in the first-principles calculation of Ref.~\onlinecite{barraza:prl10}, and they were found to induce only a negligible amount of electron scattering}
For the relevant states  in the gate voltage interval $\Delta \mu$ which do not satisfy $q\gg 1/\tilde{x}$, the adiabatic condition is satisfied for $|x| > s$ with $s =f \tilde{x}$ and  $f\gg 1$. The region $|x|>s$ is thus irrelevant for the determination of the shot noise, and we need not further consider it.~\footnote{One checks that \Eq{Esc}  implies semiclassical electron transport at $|x| > s$ not only for $-eU_0$, but also for the self-consistent potential $V(x)$, provided that $d$ stays below the critical distance $d_c$ Eq.\ (\ref{dc}).}

For screening by the conduction electrons to be negligible in our calculation of $F$, the condition $|E_{sc}| \ll E_0$ therefore needs to hold only at $|x|<s$. An evaluation of this condition requires an upper  bound on the induced electron density $n$, \Eq{n}, at $|x|<s$. Our strategy will be to obtain that density by semiclassical calculations, which are straightforward.  For many electronic states at $\varepsilon <0$, which contribute to $n$ in  \Eq{n}, the semiclassical approximation  at $|x|<s$ holds directly.  First, for all states with $|q|> f'/s$ ($f' \gg1$) the above condition for semiclassical dynamics is violated at most in an interval of length $\Delta x \ll s$ in $|x|<s$. This has a negligible effect, and we may evaluate the density $n_{lq}$ due to all states with $|q|> f'/s$ and $\varepsilon <0$ semiclassically. Similarly, the density $n_{v}$ due to all states that have $|q|<f'/s$ and $\varepsilon < -vf'/s$ may be found semiclassically.  

It then remains to find an upper bound on the charge density $n_-$ due to states $\psi_{-}$ with energies $-f' v/s<\varepsilon<0$ and $|q|<f'/s$. To this end, we apply the Friedel sum rule~\cite{friedel:phil52} at chemical potential $\varepsilon_{\rm Friedel}= vf'/s$ to all states with momenta $|q|<f'/s$. The Friedel sum rule relates scattering phase shifts  to the number of particles induced by the potential $V$. The wave functions at the energy $\varepsilon=\varepsilon_{\rm Friedel}$ and, correspondingly, the phase shifts entering the sum rule may be evaluated semiclassically. Consequently, the total particle number $N_{\rm Friedel}$ due to all states with $\varepsilon<\varepsilon_{\rm Friedel} $ and $|q|<f'/s$  can be calculated semiclassically (even though the semiclassical approximation does not hold for all involved states  individually). We then note that at $|x|\geq 2\sqrt{f'}s$ the local electron  density $n_{\rm Friedel}(x)$ due to all states with $|q|<f'/s$ at chemical potential $\varepsilon_{\rm Friedel}$ can also be obtained semiclassically. 

Since both $N_{\rm Friedel}$ and  the electron density  $n_{\rm Friedel}(x)$ for $|x|\geq 2\sqrt{f'}s$ may be found semiclassically, we conclude that also  $N_{\rm Friedel}^s$, the total  number of electrons at $|x|\leq 2\sqrt{f'}s$  for chemical potential $\varepsilon_{\rm Friedel}$ and $|q|<f'/s$ can be calculated semiclassically: $N_{\rm Friedel}^s=N_{\rm Friedel}- \int_{|x|\geq 2\sqrt{f'}s} \,dx\,n_{\rm Friedel}$.  We next decompose the electron density at $|q|<f'/s$ as $n_{\rm Friedel}= n_v+n_-+n_+$, where we also introduce  the charge density $n_+$ of all states with $0<\varepsilon<\varepsilon_{\rm Friedel}$ and $|q|<f'/s$. With this notation we have $N_{\rm Friedel}^s=\int_{-2\sqrt{f'}s}^{2\sqrt{f'}s}\,dx\,( n_v+n_{+}+n_{-})$ may be evaluated semiclassically.  Using now that the density $n_{v}$ due to all states with $\varepsilon < -vf'/s$ and $|q|<f'/s$ is semiclassical, we conclude that also $N_{+-}^s=\int_{-2\sqrt{f'}s}^{2\sqrt{f'}s}\,dx\,(n_{+}+n_{-})$ may be evaluated semiclassically. 

In order to bound  the screening field we need bounds not only on the integral of the density, but on the electron density itself. We do this by bounding its variation. Squaring \Eq{Dirac}, we have 
\beq
\left| \psi_q^\dag(x)\partial_x \psi_q(x)\right| \leq  \left| \psi_q^\dag(x) \psi_q(x)\right| \left( |\varepsilon-V(x)|/v+|q|\right)
\eeq 
for states with transverse momentum $q$ and energy $\varepsilon$. Summing over all involved $q$ and $\varepsilon$, one finds  
\beq \label{partial}
|\partial_x n_{-}(x)| \leq 4f' n_-(x)/s
\eeq
for $|x|<2\sqrt{f'}s$. We finally   integrate \Eq{partial} from $x=-2\sqrt{f'}s$ to $-s<x<s$ to find
\beqa \label{nm}
n_-(x) &\leq&  n_-(-2\sqrt{f'}s ) +(4 f'/s) \int_{-2\sqrt{f'}s}^{x} dx'\; n_-  \nn\\
&\leq& n_-(-2\sqrt{f'}s ) +(4 f'/s) \int_{-2\sqrt{f'}s}^{2\sqrt{f'}s} dx'\;( n_- +n_+) \nn\\
&\leq& n_-(-2\sqrt{f'}s ) + 4 f' N_{+-}^s /s,
\eeqa
 where we have used the non-negativity of $n_-$ and $n_+$. As shown above, both $n(-2\sqrt{f'}s )$ and $N_{+-}^s$ may be calculated semiclassically.  Equation (\ref{nm}) therefore allows us to establish  an upper bound on the total particle density $n=n_{lq}+n_v+n_-$ at $|x|<s$ and chemical potential $\varepsilon=0$  semiclassically. The calculation is straightforward and using the resulting bound on  $n$ in \Eqs{n} and (\ref{Esc}), we find that the screening field $E_{\rm sc}$ has  negligible effect, that is $|E_{sc}| \ll E_0$ at $|x|<s$, if $d \ll d_c$, where $d_c$ is the critical distance given in  \Eq{dc}.
 

\section{Appendix B} \label{appB}
In this appendix we derive the transfer matrix of electrons through the potential Eq.\ (\ref{quadpot}) in the approximations described in section \ref{analytical}.  Exploiting the inversion symmetry of the potential $V$, we may restrict our analysis to $x>0$ and obtain the transfer matrix $x<0$ by symmetry. The Hamiltonian of \Eq{Dirac} is symmetric under   $\mathcal{P}=\sigma_y\mathcal{R} $, where $\mathcal{R} $ denotes reflection on the line $x=0$, such that $\mathcal{R} \psi (x)=\psi (-x)$. Therefore, from the transfer matrix $M$ at $x,x'>0$, one obtains the one for $-x,-x'<0$ as $\mathcal{P}M\mathcal{P}^{-1}$. Then, from $M (x,0)$ for $x>0$ one finds
\beq
\label{eq:NegativeTransfer}
M(0,y)=\sigma_yM^{-1}(-y,0)\sigma_y
\eeq
at $y<0$, and accordingly the transfer matrix through the entire ribbon is
\beqa
M(x,y)&=&M(x,0)M(0,y)\nn\\
&=& M(x,0)\sigma_yM^{-1}(-y,0)\sigma_y,
\eeqa
with $x>0$ and $y<0$.

\subsubsection{Adiabatic (semiclassical) regions}
Following the work of Cheianov and Falko,~\cite{falko:prb06} the transfer matrix in the adiabatic regions, where $|q V'/V (V^2/v^2-q^2)| \ll1$, is found using the transformation
\beq
\label{eq:Ytransformation}
Y(x) = \frac{1}{V(x)}\left(\begin{array}{cc} i\chi & i\chi^* \\ V(x) & V(x)\end{array}\right),
\eeq
with $\chi=v\left[q+ik(x)\right]$ and the longitudinal wavenumber $k(x)=\sqrt{[V(x)/v]^2-q^2}$. For our form of $V$, the transverse momentum $q$ is negligible compared to $k$ throughout the adiabatic regions with $|q V'/V k^2|\ll1$, and $Y(x)$ simplifies to
\beq
Y_\pm = \left(\begin{array}{cc}-\sgn{V} & \sgn{V} \\ 1 & 1\end{array}\right),
\eeq
where $\sgn{V}$ is the sign of the potential $V(x)$. The transfer matrix for the adiabatic regions $M_{\rm ad}(x,y)$ is then given by
\beq
M_{\rm ad}(x,y)=Y(x)\tilde{M}_{\rm ad}(x,y)Y^{-1}(y),
\eeq
where the matrix $\tilde{M}_{\rm ad}(x,y)$ satisfies an equation that simplifies in the adiabatic limit $\left| q V'/V k^2\right| \ll1$ to
\beq
\label{eq:adiabaticODE}
\partial_x\tilde{M}_{\rm ad}(x,y)=i k(x)\sigma_z\tilde{M}_{\rm ad}(x,y).
\eeq

\subsubsection{Non-adiabatic regions}
The adiabatic condition $|q V'/V k^2| \ll1$ breaks down near the turning points $x=x_\pm$, where $k(x)=0$, and at $x_0$, where $V(x_0)=0$. We assume that the length of the entire non-adiabatic region at $x>0$, which includes the interval $(x_-,x_+)$, is small on the scale $x_0$ on which the potential varies, as discussed above. In this limit, we may approximate the potential $V(x)$ linearly in the non adiabatic region, 
\beq
\label{eq:linearized}
V(x)\approx 2\rho x_0 (x-x_0) = 2\sqrt{V_b \rho}(x-x_0),
\eeq
and we have $x=x_0\pm\ell$.

\begin{figure}
\centering
\includegraphics[width=0.85\columnwidth]{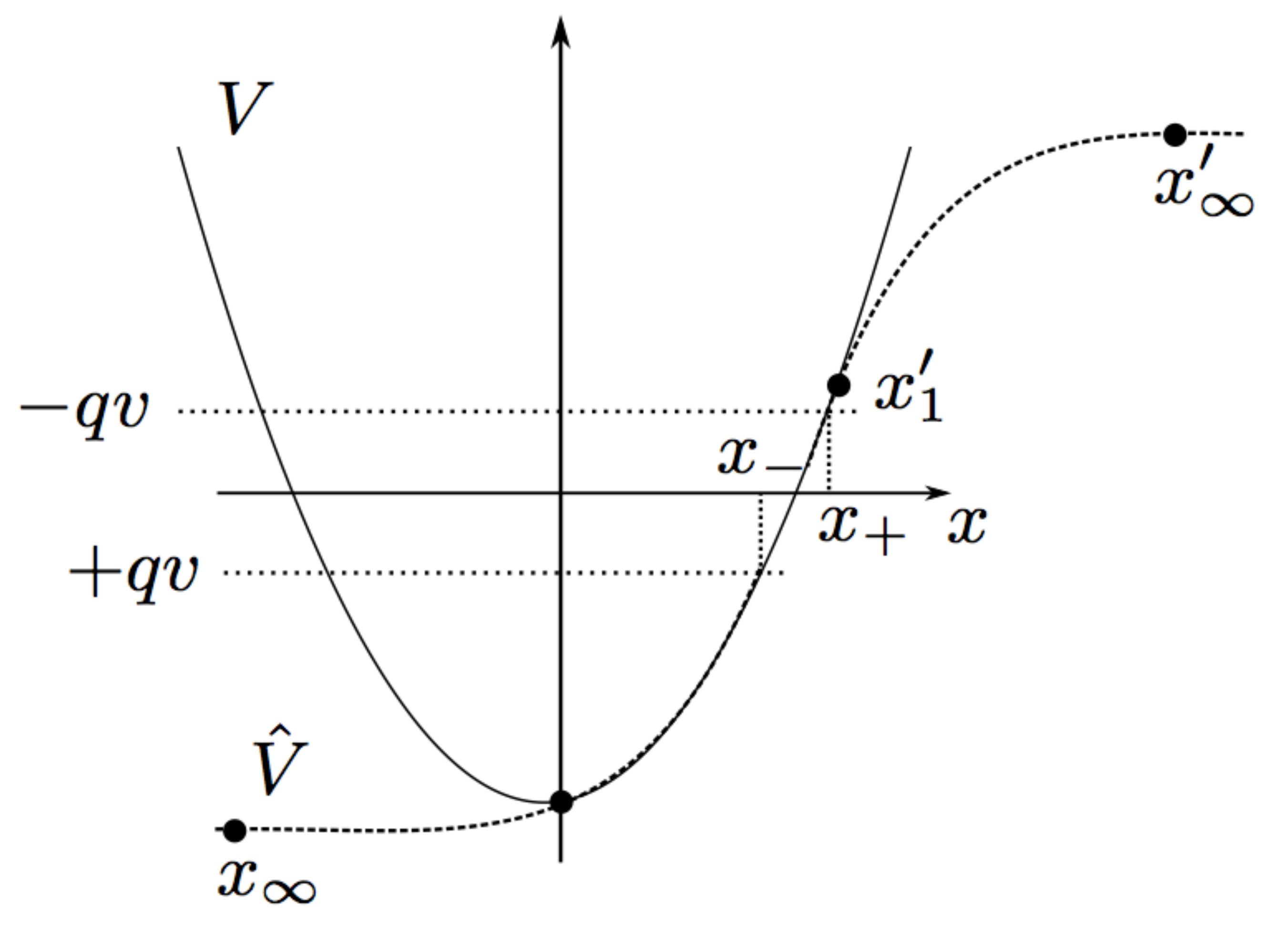}
\caption{Plots of the quadratic potential $V$ in the GNR (solid curve) and the asymptotically constant potential $\hat{V}$ considered by Cheianov and Falko~\cite{falko:prb06} (dashed curve). The two potentials coincide in the region between the points $x=0$ and $x_1'$ (here, primed coordinates refer to points to the right of the origin). We define $x_0$ such that $V(x_0)=0$. The classical turning points for the right side of the potential at $x_\pm = x_0\pm\ell$ lie within the region where $V$ can be linearized. The points $x_\infty$ and $x_\infty'$ are asymptotically far away from the turning points.} 
\label{fig:potential}
\end{figure}

Electron transport through a linear potential in graphene has an analytic solution. We use here the solution formulated in Ref.\ \onlinecite{falko:prb06} for potentials that may be linearized in the non adiabatic regions and that reach asymptotic values at $|x|\to \infty$. To this end, we define an auxiliary potential $\hat{V}$ with asymptotic values at $|x|\to\infty$ and a linear region around the turning points which coincides with the linear region of the true potential $V$, as in \Fig{fig:potential}.  Due to the assumption $|V_b| \gg v/\tilde{x}$, the potential $V$ may be linearized throughout the non adiabatic region at $x>0$, and we may choose $\hat{V}$ to coincide with $V$ in that entire region. The transfer matrix for a GNR with potential $\hat {V}$ is given by Eq.\ (\ref{eq:Mextraction}) with~\cite{falko:prb06} 
\begin{align}
\label{eq:FalkoMatrix}
\alpha &= e^{\pi\theta}, \nn \\
\beta^* &= -e^{\pi \theta/2}\frac{\sqrt{2\pi}e^{i\pi/4}\theta^{1/2+i\theta}}{\Gamma (1+i\theta)}e^{i\varphi}, \nn \\
\varphi &= k_F\ell -\int_\ell^\infty\left[ \hat{k}(x)-k_F\right]dx, \nn \\
\theta &=\frac{q^2 v}{4\sqrt{V_b\rho}},
\end{align}
where $\hat{k}(x)=\sqrt{[\hat{V}(x)/v]^2-q^2}$.

\subsubsection{Concatenation of adiabatic and non-adiabatic regions}
In order to calculate the transfer matrix for transport through the entire right side of the potential $x>0$, we combine the transfer matrices through the adiabatic regions with the solution \Eq{eq:FalkoMatrix} for transport through the auxiliary potential $\hat{V}$. Without loss of generality, we choose the auxiliary potential $\hat{V}$ to coincide with the true potential $V$ not only in the non adiabatic region, but in the entire region extending from the the potential minimum at $x=0$ up to the right end of the non-adiabatic region at $x_1'$. We obtain from \Eqs{A}, (\ref{eq:Mextraction}), and (\ref{eq:FalkoMatrix}) the transfer matrix $\hat{M}(x_\infty',x_\infty)$ of the potential $\hat{V}$, where $x_{\infty}<0$ and $x_{\infty}'>0$ are in the asymptotic regions of constant $\hat{V}$:
\beq
\hat{M}(x_\infty',x_\infty)=\hat{A}_+(x_\infty')\left(\begin{array}{cc}\alpha & \beta^* \\ \beta & \alpha^*\end{array}\right) \hat{A}_-^{-1}(x_\infty).
\eeq

To construct the transfer matrix $M(x,0)$ from $x=0$ through the non adiabatic region to a point $x>x_1'$ in the true potential $V$ (see \Fig{fig:potential}), we first perform an adiabatic transfer in $\hat{V}$ from $x=0$ to the point $x_\infty<0$. We then use $\hat{M}(x_\infty',x_\infty)$ to transport in the potential $\hat{V}$ from $x_\infty$ to a point $x_\infty'>0$. Next, we perform another adiabatic transfer in $\hat{V}$ from $x_\infty'$ back to $x_1'$, the right end of the non adiabatic region, where $\hat{V}$ and $V$ begin to deviate. The resulting transfer matrix describes transport from 0 to $x_1'$ in the potential $\hat{V}$. Finally, we transport adiabatically in the true potential $V$ from the point $x_1'$ to $x$, resulting in   
\beqa
\label{eq:PositiveTransfer}
M(x,0)=&M_{\rm ad}(x,x_1')\hat{M}_{\rm ad}(x_1',x_\infty')\times\nn\\
&\hat{M}(x_\infty',x_\infty)\hat{M}_{\rm ad}(x_\infty,0).
\eeqa
The transfer matrix $M(0,-x)$ for the left side of the potental follows from \Eqs{eq:NegativeTransfer} and (\ref{eq:PositiveTransfer}).


\begin{thebibliography}{19}
\expandafter\ifx\csname natexlab\endcsname\relax\def\natexlab#1{#1}\fi
\expandafter\ifx\csname bibnamefont\endcsname\relax
  \def\bibnamefont#1{#1}\fi
\expandafter\ifx\csname bibfnamefont\endcsname\relax
  \def\bibfnamefont#1{#1}\fi
\expandafter\ifx\csname citenamefont\endcsname\relax
  \def\citenamefont#1{#1}\fi
\expandafter\ifx\csname url\endcsname\relax
  \def\url#1{\texttt{#1}}\fi
\expandafter\ifx\csname urlprefix\endcsname\relax\def\urlprefix{URL }\fi
\providecommand{\bibinfo}[2]{#2}
\providecommand{\eprint}[2][]{\url{#2}}

\bibitem[{\citenamefont{Novoselov et~al.}(2004)\citenamefont{Novoselov, Geim,
  Morozov, Jiang, Zhang, Dubonos, Grigorieva, and Firsov}}]{novoselov:sci04}
\bibinfo{author}{\bibfnamefont{K.~S.}~\bibnamefont{Novoselov}},
  \bibinfo{author}{\bibfnamefont{A.~K.}~\bibnamefont{Geim}},
  \bibinfo{author}{\bibfnamefont{S.~V.}~\bibnamefont{Morozov}},
  \bibinfo{author}{\bibfnamefont{D.}~\bibnamefont{Jiang}},
  \bibinfo{author}{\bibfnamefont{Y.}~\bibnamefont{Zhang}},
  \bibinfo{author}{\bibfnamefont{S.~V.}~\bibnamefont{Dubonos}},
  \bibinfo{author}{\bibfnamefont{I.~V.}~\bibnamefont{Grigorieva}},
  \bibnamefont{and} \bibinfo{author}{\bibfnamefont{A.~A.}~\bibnamefont{Firsov}},
  \bibinfo{journal}{Science} \textbf{\bibinfo{volume}{306}},
  \bibinfo{pages}{666} (\bibinfo{year}{2004}).

\bibitem[{\citenamefont{Zhang et~al.}(2005)\citenamefont{Zhang, Tan, Stormer,
  and Kim}}]{zhang:nat05}
\bibinfo{author}{\bibfnamefont{Y.}~\bibnamefont{Zhang}},
  \bibinfo{author}{\bibfnamefont{Y.-W.}~\bibnamefont{Tan}},
  \bibinfo{author}{\bibfnamefont{H.~L.}~\bibnamefont{Stormer}},
  \bibnamefont{and} \bibinfo{author}{\bibfnamefont{P.}~\bibnamefont{Kim}},
  \bibinfo{journal}{Nature} \textbf{\bibinfo{volume}{438}},
  \bibinfo{pages}{201} (\bibinfo{year}{2005}).

\bibitem[{\citenamefont{Berger et~al.}(2004)\citenamefont{Berger, Song, Li, Li,
  Ogbazghi, Feng, Dai, Marchenkov, Conrad, First et~al.}}]{berger:jpc04}
\bibinfo{author}{\bibfnamefont{C.}~\bibnamefont{Berger}},
  \bibinfo{author}{\bibfnamefont{Z.}~\bibnamefont{Song}},
  \bibinfo{author}{\bibfnamefont{T.}~\bibnamefont{Li}},
  \bibinfo{author}{\bibfnamefont{X.}~\bibnamefont{Li}},
  \bibinfo{author}{\bibfnamefont{A.~Y.}~\bibnamefont{Ogbazghi}},
  \bibinfo{author}{\bibfnamefont{R.}~\bibnamefont{Feng}},
  \bibinfo{author}{\bibfnamefont{Z.}~\bibnamefont{Dai}},
  \bibinfo{author}{\bibfnamefont{A.~N.}~\bibnamefont{Marchenkov}},
  \bibinfo{author}{\bibfnamefont{E.~H.}~\bibnamefont{Conrad}},
  \bibinfo{author}{\bibfnamefont{P.~N.}~\bibnamefont{First}}, 
  \bibinfo{author}{\bibfnamefont{W.~A.}~\bibnamefont{de Heer}},
  \bibinfo{journal}{J. Phys. Chem. B}
  \textbf{\bibinfo{volume}{108}}, \bibinfo{pages}{19912}
  (\bibinfo{year}{2004}).

\bibitem[{\citenamefont{Neto et~al.}(2009)\citenamefont{Neto, Guinea, Peres,
  Novoselov, and Geim}}]{neto:rmp09}
\bibinfo{author}{\bibfnamefont{A.~H.} \bibnamefont{Castro Neto}},
  \bibinfo{author}{\bibfnamefont{F.}~\bibnamefont{Guinea}},
  \bibinfo{author}{\bibfnamefont{N.~M.~R.}~\bibnamefont{Peres}},
  \bibinfo{author}{\bibfnamefont{K.~S.}~\bibnamefont{Novoselov}},
  \bibnamefont{and} \bibinfo{author}{\bibfnamefont{A.~K.}~\bibnamefont{Geim}},
  \bibinfo{journal}{Rev. Mod. Phys.} \textbf{\bibinfo{volume}{81}},
  \bibinfo{eid}{109} (\bibinfo{year}{2009}).

\bibitem[{\citenamefont{Tworzydlo et~al.}(2006)\citenamefont{Tworzydlo,
  Trauzettel, Titov, Rycerz, and Beenakker}}]{tworzydlo:prl06}
\bibinfo{author}{\bibfnamefont{J.}~\bibnamefont{Tworzydlo}},
  \bibinfo{author}{\bibfnamefont{B.}~\bibnamefont{Trauzettel}},
  \bibinfo{author}{\bibfnamefont{M.}~\bibnamefont{Titov}},
  \bibinfo{author}{\bibfnamefont{A.}~\bibnamefont{Rycerz}}, \bibnamefont{and}
  \bibinfo{author}{\bibfnamefont{C.~W.~J.}~\bibnamefont{Beenakker}},
  \bibinfo{journal}{Phys. Rev. Lett.} \textbf{\bibinfo{volume}{96}},
  \bibinfo{eid}{246802} (\bibinfo{year}{2006}).
  
  \bibitem[{\citenamefont{DiCarlo et~al.}(2008)\citenamefont{DiCarlo, Williams,
  Zhang, McClure, and Marcus}}]{dicarlo:prl08}
\bibinfo{author}{\bibfnamefont{L.}~\bibnamefont{DiCarlo}},
  \bibinfo{author}{\bibfnamefont{J.~R.}~\bibnamefont{Williams}},
  \bibinfo{author}{\bibfnamefont{Y.}~\bibnamefont{Zhang}},
  \bibinfo{author}{\bibfnamefont{D.~T.}~\bibnamefont{McClure}},
  \bibnamefont{and} \bibinfo{author}{\bibfnamefont{C.~M.}~
  \bibnamefont{Marcus}}, \bibinfo{journal}{Phys. Rev. Lett.}
  \textbf{\bibinfo{volume}{100}}, \bibinfo{pages}{156801} (\bibinfo{year}{2008}).

\bibitem[{\citenamefont{Danneau et~al.}(2008)\citenamefont{Danneau, Wu,
  Craciun, Russo, Tomi, Salmilehto, Morpurgo, and Hakonen}}]{danneau:prl08}
\bibinfo{author}{\bibfnamefont{R.}~\bibnamefont{Danneau}},
  \bibinfo{author}{\bibfnamefont{F.}~\bibnamefont{Wu}},
  \bibinfo{author}{\bibfnamefont{M.~F.}~\bibnamefont{Craciun}},
  \bibinfo{author}{\bibfnamefont{S.}~\bibnamefont{Russo}},
  \bibinfo{author}{\bibfnamefont{M.~Y.}~\bibnamefont{Tomi}},
  \bibinfo{author}{\bibfnamefont{J.}~\bibnamefont{Salmilehto}},
  \bibinfo{author}{\bibfnamefont{A.~F.}~\bibnamefont{Morpurgo}},
  \bibnamefont{and} \bibinfo{author}{\bibfnamefont{P.~J.}~
  \bibnamefont{Hakonen}}, \bibinfo{journal}{Phys. Rev. Lett.}
  \textbf{\bibinfo{volume}{100}}, \bibinfo{pages}{196802}
  (\bibinfo{year}{2008}).
    
\bibitem[{\citenamefont{Khlus}(1987)}]{khlus:jetp87}
\bibinfo{author}{\bibfnamefont{V.~A.}~\bibnamefont{Khlus}},
  \bibinfo{journal}{Zh.  \'Eksp. Teor. Fiz.} \textbf{\bibinfo{volume}{ 93}},
  \bibinfo{pages}{2179} (\bibinfo{year}{1987}).
[\bibinfo{journal}{Sov. Phys. JETP} \textbf{\bibinfo{volume}{66}},
  \bibinfo{pages}{1243} (\bibinfo{year}{1987})].

\bibitem[{\citenamefont{Lesovik}(1989)}]{lesovik:jetp89}
\bibinfo{author}{\bibfnamefont{G.~B.}~\bibnamefont{Lesovik}},
\bibinfo{journal}{Pis'ma Zh. \'Eksp. Teor. Fiz.} \textbf{\bibinfo{volume}{49}},
  \bibinfo{pages}{513} (\bibinfo{year}{1989})
  [\bibinfo{journal}{JETP Lett.} \textbf{\bibinfo{volume}{49}},
  \bibinfo{pages}{592} (\bibinfo{year}{1989})].

\bibitem[{\citenamefont{Reznikov et~al.}(1995)\citenamefont{Reznikov, Heiblum,
  Shtrikman, and Mahalu}}]{reznikov:prl95}
\bibinfo{author}{\bibfnamefont{M.}~\bibnamefont{Reznikov}},
  \bibinfo{author}{\bibfnamefont{M.}~\bibnamefont{Heiblum}},
  \bibinfo{author}{\bibfnamefont{H.}~\bibnamefont{Shtrikman}},
  \bibnamefont{and} \bibinfo{author}{\bibfnamefont{D.}~\bibnamefont{Mahalu}},
  \bibinfo{journal}{Phys. Rev. Lett.} \textbf{\bibinfo{volume}{75}},
  \bibinfo{pages}{3340} (\bibinfo{year}{1995}).

\bibitem[{\citenamefont{Kumar et~al.}(1995)\citenamefont{Kumar, Saminadayar,
  Glatti, Jin, and Etienne}}]{kumar:prl96}
\bibinfo{author}{\bibfnamefont{A.}~\bibnamefont{Kumar}},
  \bibinfo{author}{\bibfnamefont{L.}~\bibnamefont{Saminadayar}},
  \bibinfo{author}{\bibfnamefont{D.~C.}~\bibnamefont{Glattli}}, 
  \bibinfo{author}{\bibfnamefont{Y.}~\bibnamefont{Jin}},
  \bibnamefont{and} \bibinfo{author}{\bibfnamefont{B.}~\bibnamefont{Etienne}},
  \bibinfo{journal}{Phys. Rev. Lett.} \textbf{\bibinfo{volume}{76}},
  \bibinfo{pages}{2778} (\bibinfo{year}{1996}).
  
  \bibitem[{\citenamefont{Cayssol et~al.}(2009)\citenamefont{Cayssol, Huard,
  and Goldhaber-Gordon}}]{cayssol:prb09}
\bibinfo{author}{\bibfnamefont{J.}~\bibnamefont{Cayssol}},
  \bibinfo{author}{\bibfnamefont{B.}~\bibnamefont{Huard}},
  \bibnamefont{and} \bibinfo{author}{\bibfnamefont{D.}~
  \bibnamefont{Goldhaber-Gordon}}, \bibinfo{journal}{Phys. Rev. B} \textbf{\bibinfo{volume}{79}},
  \bibinfo{pages}{075428} (\bibinfo{year}{2009}).

\bibitem[{\citenamefont{Rycerz et~al.}(2009)\citenamefont{Rycerz,
  Recher, and Wimmer}}]{rycerz:prb09}
\bibinfo{author}{\bibfnamefont{A.}~\bibnamefont{Rycerz}},
  \bibinfo{author}{\bibfnamefont{P.}~\bibnamefont{Recher}},
  \bibnamefont{and} \bibinfo{author}{\bibfnamefont{M.}~\bibnamefont{Wimmer}},
  \bibinfo{journal}{Phys. Rev. B} \textbf{\bibinfo{volume}{80}},
  \bibinfo{eid}{125417} (\bibinfo{year}{2009}).

\bibitem[{\citenamefont{Snyman et~al.}(2007)\citenamefont{Snyhman,
  and Beenakker}}]{snyman:prb07}
\bibinfo{author}{\bibfnamefont{I.}~\bibnamefont{Snyman}}
  \bibnamefont{and} \bibinfo{author}{\bibfnamefont{C.~W.~J.}~\bibnamefont{Beenakker}},
  \bibinfo{journal}{Phys. Rev. B} \textbf{\bibinfo{volume}{75}},
  \bibinfo{eid}{045322} (\bibinfo{year}{2007}).

\bibitem[{\citenamefont{San-Jose et~al.}(2007)\citenamefont{San-Jose, Prada,
  and Golubev}}]{sanjose:prb07}
\bibinfo{author}{\bibfnamefont{P.}~\bibnamefont{San-Jose}},
  \bibinfo{author}{\bibfnamefont{E.}~\bibnamefont{Prada}}, \bibnamefont{and}
  \bibinfo{author}{\bibfnamefont{D.~S.}~\bibnamefont{Golubev}},
  \bibinfo{journal}{Phys. Rev. B} \textbf{\bibinfo{volume}{76}},
  \bibinfo{pages}{195445} (\bibinfo{year}{2007}).

\bibitem[{\citenamefont{Lewenkopf et~al.}(2008)\citenamefont{Lewenkopf,
  Mucciolo, and Castro~Neto}}]{lewenkopf:prb08}
\bibinfo{author}{\bibfnamefont{C.~H.}~\bibnamefont{Lewenkopf}},
  \bibinfo{author}{\bibfnamefont{E.~R.}~\bibnamefont{Mucciolo}},
  \bibnamefont{and} \bibinfo{author}{\bibfnamefont{A.~H.}~
  \bibnamefont{Castro~Neto}}, \bibinfo{journal}{Phys. Rev. B}
  \textbf{\bibinfo{volume}{77}}, \bibinfo{pages}{081410(R)}
  (\bibinfo{year}{2008}).
  
\bibitem[{\citenamefont{Schuessler et~al.}(2009)\citenamefont{Schuessler,
  Ostrovsky, Gornyi, and Mirlin}}]{schuessler:prb09}
\bibinfo{author}{\bibfnamefont{A.}~\bibnamefont{Schuessler}},
  \bibinfo{author}{\bibfnamefont{P.~M.}~\bibnamefont{Ostrovsky}},
  \bibinfo{author}{\bibfnamefont{I.~V.}~\bibnamefont{Gornyi}},
  \bibnamefont{and} \bibinfo{author}{\bibfnamefont{A.~D.}~
  \bibnamefont{Mirlin}}, \bibinfo{journal}{Phys. Rev. B}
  \textbf{\bibinfo{volume}{79}}, \bibinfo{pages}{075405}
  (\bibinfo{year}{2009}).

\bibitem[{\citenamefont{Sonin}(2008)}]{sonin:prb08}
\bibinfo{author}{\bibfnamefont{E.~B.}~\bibnamefont{Sonin}},
  \bibinfo{journal}{Phys. Rev. B} \textbf{\bibinfo{volume}{77}},
  \bibinfo{pages}{233408} (\bibinfo{year}{2008}).

\bibitem[{\citenamefont{Golub and Horovitz}(2010)}]{golub:prb10}
\bibinfo{author}{\bibfnamefont{A.}~\bibnamefont{Golub}} \bibnamefont{and}
  \bibinfo{author}{\bibfnamefont{B.}~\bibnamefont{Horovitz}},
  \bibinfo{journal}{Phys. Rev. B} \textbf{\bibinfo{volume}{81}},
  \bibinfo{pages}{245424} (\bibinfo{year}{2010}).

\bibitem[{\citenamefont{Wiener et~al.}(2009)\citenamefont{Wiener, and Kindermann}}]{wiener:prb11}
\bibinfo{author}{\bibfnamefont{A.~D.}~\bibnamefont{Wiener}} \bibnamefont{and}
  \bibinfo{author}{\bibfnamefont{M.}~\bibnamefont{Kindermann}},
  \bibinfo{journal}{Phys. Rev. B} \textbf{\bibinfo{volume}{84}},
  \bibinfo{pages}{245420} (\bibinfo{year}{2011}).

\bibitem[{\citenamefont{Barraza-Lopez et~al.}(2010)\citenamefont{Barraza-Lopez, Vanevic,
  Kindermann, and Chou}}]{barraza:prl10}
\bibinfo{author}{\bibfnamefont{S.}~\bibnamefont{Barraza-Lopez}},
  \bibinfo{author}{\bibfnamefont{M.}~\bibnamefont{Vanevi\'c}},
  \bibinfo{author}{\bibfnamefont{M.}~\bibnamefont{Kindermann}},
  \bibnamefont{and} \bibinfo{author}{\bibfnamefont{M.~Y.}~\bibnamefont{Chou}},
  \bibinfo{journal}{Phys. Rev. Lett.} \textbf{\bibinfo{volume}{104}},
  \bibinfo{eid}{076807} (\bibinfo{year}{2010}).

\bibitem[{\citenamefont{Barraza-Lopez et~al.}(2012)\citenamefont{Barraza-Lopez,
  Kindermann, and Chou}}]{barraza:nano12}
\bibinfo{author}{\bibfnamefont{S.}~\bibnamefont{Barraza-{L}opez}},
  \bibinfo{author}{\bibfnamefont{M.}~\bibnamefont{Kindermann}},
  \bibnamefont{and} \bibinfo{author}{\bibfnamefont{M.~Y.}~\bibnamefont{Chou}},
  \bibinfo{journal}{Nano Lett.} \textbf{\bibinfo{volume}{12}},
  \bibinfo{eid}{3424} (\bibinfo{year}{2012}).
  
  \bibitem[{\citenamefont{Berry et~al.}(1987)\citenamefont{Berry, and Mondragon}}]{berry:prsla87}
\bibinfo{author}{\bibfnamefont{M. V.}~\bibnamefont{Berry}},
  \bibnamefont{and} \bibinfo{author}{\bibfnamefont{R. J.}~\bibnamefont{Mondragon}},
  \bibinfo{journal}{Proc. R. Soc. Lond.} \textbf{\bibinfo{volume}{A 412}},
  \bibinfo{pages}{53} (\bibinfo{year}{1987}).
  
   \bibitem[{\citenamefont{Brey et~al.}(2006)\citenamefont{Brey, and Fertig}}]{brey:prb06}
\bibinfo{author}{\bibfnamefont{L.}~\bibnamefont{Brey}},
  \bibnamefont{and} \bibinfo{author}{\bibfnamefont{H. A.}~\bibnamefont{Fertig}},
  \bibinfo{journal}{Phys. Rev. B} \textbf{\bibinfo{volume}{73}},
  \bibinfo{eid}{235411} (\bibinfo{year}{2006}).
  
   \bibitem[{\citenamefont{Gray et~al.}(2009)\citenamefont{Gray, Hurst, Lewis, and Meyer}}]{gray:09}
  \bibinfo{author}{\bibfnamefont{P.~R.}~\bibnamefont{Gray}}, 
   \bibinfo{author}{\bibfnamefont{P.~J.}~\bibnamefont{Hurst}},
   \bibinfo{author}{\bibfnamefont{S.~H.}~\bibnamefont{Lewis}},
   \bibnamefont{and} \bibinfo{author}{\bibfnamefont{R.~G.}~\bibnamefont{Meyer}},
   \textit{\bibinfo{title}{Analysis and design of analog integrated circuits, fifth ed}}.
   \bibinfo{publisher}{Wiley}(\bibinfo{year}{2009}).
  
  \bibitem[{\citenamefont{Cheianov et~al.}(2009)\citenamefont{Cheianov, and Fal'ko}}]{falko:prb06}
\bibinfo{author}{\bibfnamefont{V.~V.}~\bibnamefont{Cheianov}} \bibnamefont{and}
  \bibinfo{author}{\bibfnamefont{V.~I.}~\bibnamefont{Fal'ko}},
  \bibinfo{journal}{Phys. Rev. B} \textbf{\bibinfo{volume}{74}},
  \bibinfo{pages}{041403(R)} (\bibinfo{year}{2006}).
  
  
\bibitem[{\citenamefont{B\"uttiker}(1990)}]{buettiker:prl90}
\bibinfo{author}{\bibfnamefont{M.}~\bibnamefont{B\"uttiker}},
  \bibinfo{journal}{Phys. Rev. Lett.} \textbf{\bibinfo{volume}{65}},
  \bibinfo{pages}{2901} (\bibinfo{year}{1990}).
  
  \bibitem[{\citenamefont{Friedel}(1952)}]{friedel:phil52}
\bibinfo{author}{\bibfnamefont{J.}~\bibnamefont{Friedel}},
  \bibinfo{journal}{Philos. Mag.} \textbf{\bibinfo{volume}{43}},
  \bibinfo{pages}{153} (\bibinfo{year}{1952}).  
    
 %
%
%
%
%
%
%
%
%
%
%
%
%
%
%
  
\end{thebibliography}
\end{document}